\documentclass[conference,letterpaper]{IEEEtran}
\IEEEoverridecommandlockouts

\usepackage[letterpaper, top=0.75in, bottom=1.08in, left=0.65in, right=0.65in]{geometry}

\usepackage{cite}
\usepackage{bm}
\usepackage{amsmath,amssymb,amsfonts}
\usepackage{algorithmic}
\usepackage{graphicx}
\usepackage{textcomp}
\usepackage{xcolor}

\usepackage{enumitem}
\setlist{nosep}

\allowdisplaybreaks

\def\BibTeX{{\rm B\kern-.05em{\sc i\kern-.025em b}\kern-.08em
    T\kern-.1667em\lower.7ex\hbox{E}\kern-.125emX}}

\newtheorem{theorem}{Theorem}
\newtheorem{definition}{Definition}
\newtheorem{lemma}{Lemma}
\newtheorem{assumption}{Assumption}
\newtheorem{remark}{Remark}

\begin{document}

\title{Sensing-Limited Control of Noiseless Linear Systems Under Nonlinear Observations
\thanks{Corresponding author: Fan Liu.}
}

\author{\IEEEauthorblockN{Ming Li$^1$, Fan Liu$^{2}$, Yifeng Xiong$^3$, Jie Xu$^4$, Tao Liu$^1$}
\IEEEauthorblockA{
\textit{$^{1}$Southern University of Science and Technology, Shenzhen, China}\\
\textit{$^{2}$Southeast University, Nanjing, China}\\
\textit{$^3$Beijing University of Posts and Telecommunications, Beijing, China}\\
\textit{$^4$The Chinese University of Hong Kong, Shenzhen, China}\\
E-mail: lim2024@mail.sustech.edu.cn; fan.liu@seu.edu.cn; \\yifengxiong@bupt.edu.cn; xujie@cuhk.edu.cn; liut6@sustech.edu.cn}
}

\maketitle

\begin{abstract}
This paper investigates the fundamental information-theoretic limits for the control and sensing of noiseless linear dynamical systems subject to a broad class of nonlinear observations. We analyze the interactions between the control and sensing components by characterizing the minimum information flow required for stability. Specifically, we derive necessary conditions for  mean-square observability and stabilizability, demonstrating that the average directed information rate from the state to the observations must exceed the intrinsic expansion rate of the unstable dynamics. Furthermore, to address the challenges posed by non-Gaussian distributions inherent to nonlinear observation channels, we establish sufficient conditions by imposing regularity assumptions, specifically log-concavity, on the system's probabilistic components. We show that under these conditions, the divergence of differential entropy implies the convergence of the estimation error, thereby closing the gap between information-theoretic bounds and estimation performance. By establishing these results, we unveil the fundamental performance limits imposed by the sensing layer, extending classical data-rate constraints to the more challenging regime of nonlinear observation models.
\end{abstract}

\begin{IEEEkeywords}
Information theory, sensing, feedback control, nonlinear observations.
\end{IEEEkeywords}

\section{Introduction}
Control under communication constraints constitutes a foundational pillar of modern systems theory, driven by the necessity to operate dynamical systems over digital networks or analog links with finite capacity \cite{liberzon2003stabilization, tatikonda2004control}. A central theme of this literature has been the characterization of stabilizability limits for unstable plants subject to data-rate restrictions. Classical theory establishes a fundamental limit for stabilization: the feedback loop's data rate $\mathcal{C}$ must exceed a critical threshold $\mathcal{R}$\cite{nair2007feedback,tatikonda2004controlnoisy,tatikonda2004control,sahai2001anytime,Girish2004rates}. This quantity $\mathcal{R}$, determined by the unstable eigenvalues of the system matrix, represents the intrinsic entropy generation rate of the dynamics. Crucially, the precise nature of these stabilization conditions depends on the stability criterion employed, which is broadly classified into two categories: stability in the almost sure sense \cite{tatikonda2004controlnoisy,tatikonda2004control} and moment stability \cite{sahai2000evaluating, sahai2001anytime,Girish2004rates}.

While communication constraints remain theoretically significant, the exponential growth of bandwidth in modern infrastructure has rendered pure transmission limits less critical in many applications. Instead, in emerging domains such as humanoid robotics \cite{sun2025learning,peng2021amp} and autonomous driving \cite{martinez2018autonomous}, the bottleneck has decisively shifted from the communication channel to the \textit{sensing channel}. As these systems navigate complex, unstructured environments, they are flooded with high-dimensional sensory data. In this regime, performance is circumscribed not by the capacity to transmit bits, but by the efficiency of perceiving and extracting task-relevant information from the raw observation stream \cite{gunduz2022beyond}. Consequently, the sensing mechanism has emerged as the pivotal constraint in the closed-loop architecture, yet its information-theoretic implications on control remain underexplored relative to their communication counterparts.

A fundamental theoretical gap persists: we still lack a rigorous understanding of the performance limits imposed strictly by sensing. Crucially, existing theorems from communication-limited control cannot be trivially extrapolated to the sensing domain due to a fundamental structural difference regarding the encoding process. In classical networked control, the encoder is typically a computational module that can be arbitrarily designed to optimize information transmission \cite{tatikonda2004control}. This design freedom allows for sophisticated coding schemes that match the source statistics to the channel capacity. In contrast, the ``encoder'' in a sensing problem is the physical observation mechanism itself, which is governed by immutable physical laws rather than algorithmic design.  Standard data-rate theorems, which implicitly assume optimal encoding capability, thus fail to capture the constraints imposed by the rigid geometry of sensing channels. Without a unified framework to analyze these inherent sensing limitations, system design often resorts to heuristics, leaving open the fundamental question: what is the minimum sensory information requisite for achieving a specific control objective? Deriving these theoretical bounds is essential for transcending empirical methods and providing a principled basis for sensing-aware control design.

\begin{figure*}[t]
    \centering
    \includegraphics[width=0.7\linewidth]{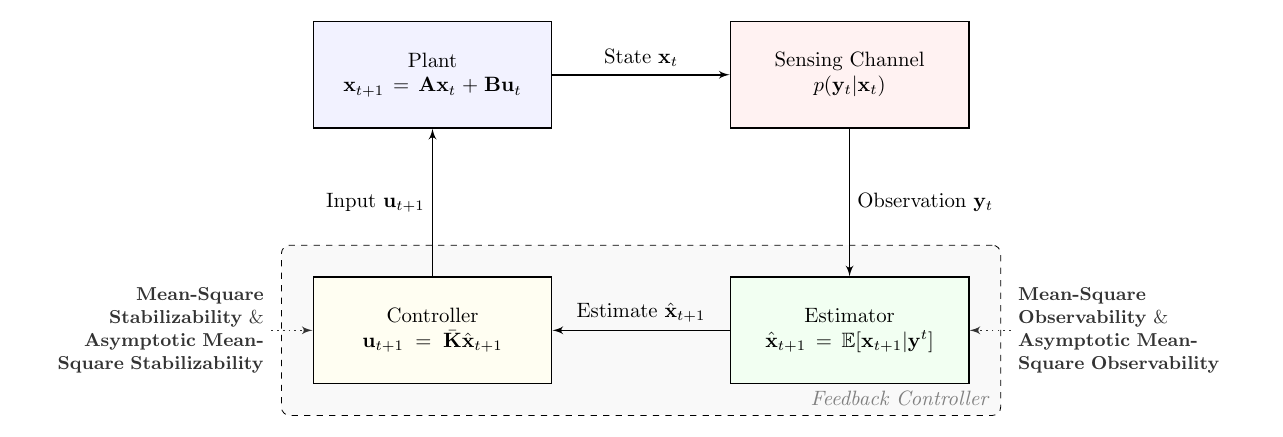} 
    \caption{Closed-loop control system with nonlinear sensing.}
    \label{fig:system_model}
\end{figure*}

This work investigates the sensing-control nexus from an information-theoretic perspective. We leverage the concept of directed information to explicitly quantify the causal information flow required for stabilization. We establish necessary conditions for both mean-square observability and stabilizability, proving that the average directed information flow from the unstable state to the observations must, at a minimum, counterbalance the intrinsic expansion rate of the system dynamics. To bridge the gap between theoretical bounds and achievability, we further derive sufficient conditions for asymptotic observability and stabilizability. We demonstrate that under specific regularity assumptions, a directed information rate strictly exceeding the expansion rate guarantees successful estimation and stabilization. Collectively, these results characterize the fundamental performance limits imposed by the sensing layer, extending classical information constraints to the challenging regime of a broad class of nonlinear sensing channels.

\section{System Model}

\subsection{Dynamics and Sensing}
Consider the discrete-time linear system (Fig. \ref{fig:system_model}):
\begin{equation}
    \mathbf{x}_{t+1} = \mathbf{A} \mathbf{x}_t + \mathbf{B} \mathbf{u}_t, \quad t \in \mathbb{N}_0,
    \label{eq:system_dynamics}
\end{equation}
where ${\bf x}_t \in \mathbb{R}^n$ is the state vector, ${\bf u}_t \in \mathbb{R}^m$ is the control input vector, ${\bf A} \in \mathbb{R}^{n \times n}$ is the system matrix, and ${\bf B} \in \mathbb{R}^{n \times m}$ is the control matrix. We assume that the system matrix ${\bf A}$ is unstable (i.e., it has at least one eigenvalue with magnitude greater than or equal to 1) and that the initial state ${\bf x}_0$ has a valid probability density function (PDF) with a finite second moment. 

The system is observed through a memoryless vector channel subject to noise and nonlinear distortions, yielding the observation ${\bf y}_t \in \mathbb{R}^p$. To capture these stochastic and nonlinear characteristics, the observation process is modeled by a general conditional PDF $p({\bf y}_t|{\bf x}_t)$. We denote the sequence of observations up to time $t$ as ${\bf y}^t = \{{\bf y}_0, {\bf y}_1, \dots, {\bf y}_t\}$.

\subsection{Canonical Decomposition}
To isolate the information-critical dynamics, we decouple the system into stable and unstable modes via a Jordan decomposotion $\mathbf{z}_t = \mathbf{T}\mathbf{x}_t = [(\mathbf{z}_t^u)^T, (\mathbf{z}_t^s)^T]^T$. This yields the block-diagonal forms $\mathbf{A}_z = \text{diag}(\mathbf{A}_u, \mathbf{A}_s)$ and $\mathbf{B}_z = [\mathbf{B}_u^T, \mathbf{B}_s^T]^T$. Since $\mathbf{A}_s$ is Schur stable\footnote{A matrix is Schur stable if all its eigenvalues strictly lie inside the unit circle, i.e., $|\lambda_i| < 1$.}, Input-to-State Stability (ISS) implies that stabilizing the unstable subspace $\mathbf{z}_t^u$ suffices for global stability. The information-theoretic limit is captured by the \textit{expansion rate}:
\begin{equation}
   R_{\mathrm{exp}} := \sum_{|\lambda_i(\mathbf{A})| \ge 1} \log_2 |\lambda_i(\mathbf{A})| = \log_2 |\det(\mathbf{A}_u)|.
\end{equation}

\subsection{Definitions of Stabilizability and Observability}
We distinguish between \textit{boundedness} (used for necessary conditions) and \textit{asymptotic convergence} (used for sufficient conditions).

\begin{definition}[Mean-Square Stabilizability] \label{def:ms_stab}
    The system is said to be \textit{mean-square stabilizable} if there exists a control strategy ${\bf u}_t$ such that the second moment of the state remains bounded:
    \begin{equation}
        \limsup_{t \to \infty} \mathbb{E}[\|{\bf x}_t\|^2] < \infty.
    \end{equation}
\end{definition}

\begin{definition}[Asymptotic Mean-Square Stabilizability] \label{def:asymp_ms_stab}
    The system is said to be \textit{asymptotically mean-square stabilizable} if there exists a control strategy ${\bf u}_t$ such that for any initial state ${\bf x}_0$, the following two properties are satisfied:
    \begin{enumerate}
        \item Stability: $\forall \epsilon > 0, \exists \delta(\epsilon) > 0$ such that if $\mathbb{E}[\|{\bf x}_0\|^2] \le \delta$, then $\mathbb{E}[\|{\bf x}_t\|^2] \le \epsilon$ for all $t \in \mathbb{N}_0$.
        \item Attractivity: $\lim_{t \to \infty} \mathbb{E}[\|{\bf x}_t\|^2] = 0$.
    \end{enumerate}
\end{definition}

\begin{definition}[Mean-Square Observability] \label{def:ms_obs}
    The system is said to be \textit{mean-square observable} if there exists an estimation strategy providing an estimate $\hat{{\bf x}}_t$ such that  the second moment of the estimation error ${\bf e}_t := \hat{{\bf x}}_t - {\bf x}_t$ remains bounded in the mean-square sense:
    \begin{equation}
        \limsup_{t \to \infty} \mathbb{E}[\|{\bf e}_t\|^2] < \infty.
    \end{equation}
\end{definition}

\begin{definition}[Asymptotic Mean-Square Observability] \label{def:asymp_ms_obs}
    The system is said to be \textit{asymptotically mean-square observable} if there exists an estimation strategy such that for any initial error ${\bf e}_0$, the following two properties are satisfied:
    \begin{enumerate}
        \item Stability: $\forall \epsilon > 0, \exists \delta(\epsilon) > 0$ such that if $\mathbb{E}[\|{\bf e}_0\|^2] \le \delta$, then $\mathbb{E}[\|{\bf e}_t\|^2] \le \epsilon$ for all $t \in \mathbb{N}_0$.
        \item Attractivity: $\lim_{t \to \infty} \mathbb{E}[\|{\bf e}_t\|^2] = 0$.
    \end{enumerate}
\end{definition}

\begin{remark}
    Asymptotic stability (Definitions  \ref{def:asymp_ms_stab} and \ref{def:asymp_ms_obs}) strictly implies mean-square boundedness (Definitions \ref{def:ms_stab} and \ref{def:ms_obs}). Consequently, our necessary conditions target the weaker boundedness notion to maximize generality, while sufficiency proofs demonstrate the stronger asymptotic convergence under sufficient information flow.
\end{remark}

\section{Necessary Conditions}
To quantify the causal information flow required for stabilization, we leverage the concept of \textit{directed information}. Unlike mutual information, which captures symmetric correlation, directed information explicitly measures the information transfer from the state process to the observation sequence over the feedback channel. Specifically, we define the cumulative directed information from the unstable state sequence to the observations up to time $T$ as:
\begin{equation}
    I(\mathbf{z}_T^u \to \mathbf{y}^T) := \sum_{t=0}^T I(\mathbf{z}_t^u; \mathbf{y}_t | \mathbf{y}^{t-1}).
\end{equation}
This metric characterizes the cumulative reduction in uncertainty about the unstable states provided by the observation history.

\textbf{Remark 1:} 
Compared to standard mutual information $I(\mathbf{x}^T; \mathbf{y}^T)$, the directed information $I(\mathbf{x}^T \to \mathbf{y}^T)$ offers two critical advantages for sensing-limited control. First, it strictly isolates ``fresh'' sensory information by sequentially conditioning on observation history, which causally decouples the deterministic feedback control from the stochastic observation process. Second, it provides a clear meaning for perception, explicitly quantifying the rate at which the sensing mechanism eliminates the intrinsic entropy generated by the unstable dynamics.

We now establish that the sensing channel must drain entropy at least as fast as the unstable dynamics generate it.

\begin{theorem}[Necessity for Observability] \label{thm-Necessary-Observability}
    If the system is mean-square observable, the average directed information rate satisfies:
    \begin{equation}
        \liminf_{T \to \infty} \frac{1}{T+1} I(\mathbf{z}_T^u \to \mathbf{y}^T) \ge R_{\mathrm{exp}}.
    \end{equation}
\end{theorem}

\begin{IEEEproof}
    Let $h_t := h(\mathbf{z}_t^u | \mathbf{y}^{t-1})$. The entropy evolution under unstable dynamics $\mathbf{z}_{t+1}^u = \mathbf{A}_u \mathbf{z}_t^u + \mathbf{B}_u \mathbf{u}_t$ is:
\begin{align}
    h(\mathbf{z}_{t+1}^u | \mathbf{y}^t) 
    &= h(\mathbf{A}_u \mathbf{z}_t^u + \mathbf{B}_u \mathbf{u}_t | \mathbf{y}^t) \nonumber \\
    &= h(\mathbf{A}_u \mathbf{z}_t^u | \mathbf{y}^t) \label{eq:entropy_step1} \\
    &= h(\mathbf{z}_t^u | \mathbf{y}^t) + \log_2 |\det \mathbf{A}_u| \nonumber \\
    &= h(\mathbf{z}_t^u | \mathbf{y}^t) +R_{\mathrm{exp}}, \label{eq:entropy_evolution}
\end{align}
where \eqref{eq:entropy_step1} holds because $\mathbf{u}_t$ is a deterministic function of $\mathbf{y}^{t-1}$, and thus $\mathbf{y}^t$, representing a coordinate shift that does not alter differential entropy.
    Using $h(\mathbf{z}_t^u | \mathbf{y}^t) = h_t - I(\mathbf{z}_t^u; \mathbf{y}_t | \mathbf{y}^{t-1})$, we get the recursion:
    \begin{equation}
        h_{t+1} = h_t +R_{\mathrm{exp}} - I(\mathbf{z}_t^u; \mathbf{y}_t | \mathbf{y}^{t-1}). \notag
    \end{equation}
    Summing from $0$ to $T$ and dividing by $T+1$:
    \begin{equation} \label{eq:rate_balance}
        \frac{1}{T+1} I(\mathbf{z}_T^u \to \mathbf{y}^T) =R_{\mathrm{exp}} + \frac{h_0 - h_{T+1}}{T+1}. \notag
    \end{equation}
    \enlargethispage{-0.3in}
    Mean-square Observability implies $\mathbb{E}[\|\mathbf{e}_t^u\|^2] \le C$. Since Gaussian distributions maximize entropy for a fixed covariance, the conditional entropy is bounded:
    \begin{align}
        h_{T+1} &= h(\mathbf{e}_{T+1}^u | \mathbf{y}^T) \le h(\mathbf{e}_{T+1}^u) \le \frac{n_u}{2} \log_2 \left( \frac{2\pi e}{n_u} \text{Tr}(\mathbf{\Sigma}_{e}) \right) \nonumber \\
        &\le \mathcal{H}_{\mathrm{max}}:=\frac{n_u}{2} \log_2 \left( \frac{2\pi e}{n_u} C \right), \notag
    \end{align}
    where $\mathcal{H}_{\mathrm{max}}$ is constant derived from the uniform error bound $C$. Since both $\mathcal{H}_{\mathrm{max}}$ and the initial entropy $h_0$ are finite constants independent of $T$, the result follows.
\end{IEEEproof}

\begin{theorem}[Necessity for Stabilizability]
    If the system is mean-square stabilizable, then $\liminf_{T \to \infty} \frac{1}{T+1} I(\mathbf{z}_T^u \to \mathbf{y}^T) \ge R_{\mathrm{exp}}$.
\end{theorem}
\begin{IEEEproof}
    The proof mirrors Theorem \ref{thm-Necessary-Observability}. Mean-square Stabilizability implies $\mathbb{E}[\|\mathbf{z}_t^u\|^2] \le C_{\mathrm{max}}$. Thus, the terminal entropy $h(\mathbf{z}_{T+1}^u | \mathbf{y}^T)$ is upper-bounded by a constant $\mathcal{H}_{\mathrm{bound}}$. Taking the limit in \eqref{eq:rate_balance} yields the inequality.
\end{IEEEproof}

\textbf{Remark 2:} 
The key distinction between our work and the classical results in \cite{tatikonda2004control} lies in the paradigm shift regarding the encoding process. In the communication-constrained framework of \cite{tatikonda2004control}, the encoder is a computational module (i.e., a digital quantizer) that can be arbitrarily designed to match the available channel rate. In contrast, our innovation focuses on the \textit{sensing-limited} regime, where the ``encoder'' is defined by the physical observation mechanism $p({\bf y}_t|{\bf x}_t)$ itself. This mechanism is governed by immutable physical laws rather than algorithmic design. By extending classical data-rate theorems to the sensing domain, we characterize fundamental performance boundaries imposed by the sensing layer that cannot be circumvented by algorithmic optimization.

\section{Sufficient Conditions}

Sufficiency requires bridging the gap between entropy and variance. While $h(\mathbf{x}) \to -\infty$ implies $\det(\mathbf{\Sigma}) \to 0$ for Gaussians, this is not universal (e.g., multimodal distributions). To preclude pathological cases, we impose regularity constraints.

\subsection{Assumptions}

\begin{assumption} \label{assump:obs_hessian}
    The observation log-likelihood is $C^2$. There exist $L \ge 1, \alpha > 0$ such that the cumulative Hessian satisfies:
    \begin{equation}
        \sum_{k=t-L+1}^t \nabla_{\mathbf{z}_t^u}^2 \log p(\mathbf{y}_k | \mathbf{z}_k^u) \preceq -\alpha \mathbf{I}, \quad \forall t \ge L-1. \notag
    \end{equation}
    where past states depend on $\mathbf{z}_t^u$ via inverse dynamics.
\end{assumption}

\begin{assumption} \label{assump:prior_hessian}
The initial unstable state $\mathbf{z}_0^u$ follows a PDF $p(\mathbf{z}_0^u)$ that is twice continuously differentiable. There exists a scalar $\beta > 0$ such that for all $\mathbf{z}_0^u \in \mathbb{R}^{n_u}$:
\begin{equation}
    \nabla_{\mathbf{z}_0^u}^2 \log p(\mathbf{z}_0^u) \preceq \beta \cdot \mathbf{I}. \notag
    \label{eq:assump_prior_hessian}
\end{equation}
\end{assumption}

\begin{assumption} \label{assump:cond_number}
The condition number of the conditional estimation error covariance matrix $\mathbf{\Sigma}_{t|t} = \mathbb{E}[(\mathbf{x}_t - \hat{\mathbf{x}}_t)(\mathbf{x}_t - \hat{\mathbf{x}}_t)^T | \mathbf{y}^t]$ is uniformly bounded:
\begin{equation}
    \frac{\lambda_{\max}(\mathbf{\Sigma}_{t|t})}{\lambda_{\min}(\mathbf{\Sigma}_{t|t})} \le \kappa < \infty, \quad \forall t \in \mathbb{N}_0,\notag
    \label{eq:assump_cond_number}
\end{equation}
where $\lambda_{\max}(\cdot)$ and $\lambda_{\min}(\cdot)$ denote the maximum and minimum eigenvalues, respectively.
\end{assumption}

\subsection{Theorems}

We utilize the following property of log-concave densities: 
Let ${\bf x} \in \mathbb{R}^n$ be a random vector with an absolutely continuous log-concave density. Let ${\bf z} \sim \mathcal{N}(\bm{\mu}, {\bf \Sigma})$ be a Gaussian random vector with the same covariance matrix ${\bf \Sigma} = \mathbb{E}[({\bf x}-\bm{\mu})({\bf x}-\bm{\mu})^T]$. Then, the differential entropy of ${\bf x}$ satisfies the following bounds relative to the Gaussian entropy:
    \begin{equation} \label{eq:entropy_gap}
        0 \le \frac{1}{n} h({\bf z}) - \frac{1}{n} h({\bf x}) \le C,
    \end{equation}
    where $C$ is a universal constant.

\begin{IEEEproof}
    The bounds in \eqref{eq:entropy_gap} are established in \cite{Bobkov2011} and proved in \cite[Corollary 4.2]{Bobkov2012}. 
\end{IEEEproof}

    Consequently, the entropy of a log-concave vector is fundamentally tied to its covariance determinant:
    $h(\mathbf{x}) \to -\infty \iff \det(\mathbf{\Sigma}) \to 0$.
    
\begin{theorem}[Sufficiency for Asymptotic Observability]\label{thm-Sufficiency-Observability}
    Under Assumptions 1--3, if $\liminf_{T \to \infty} \frac{1}{T+1} I(\mathbf{z}_T^u \to \mathbf{y}^T) >R_{\mathrm{exp}}$, the system is asymptotically mean-square observable.
\end{theorem}

\begin{IEEEproof}
    We begin by analyzing the evolution of the conditional differential entropy of the unstable state ${\bf z}_t^u$. Let $h(\cdot | {\bf y}^{t-1})$ and $h(\cdot | {\bf y}^t)$ denote the prior and posterior entropies, respectively. The entropy satisfies the following recursion:
    \begin{align} \label{eq:entropy_recursion}
        h({\bf z}_{t+1}^u | {\bf y}^t) 
        &= h({\bf A}_u {\bf z}_t^u + {\bf B}_u {\bf u}_t | {\bf y}^t) \notag \\
        &\stackrel{(a)}{=} h({\bf z}_t^u | {\bf y}^t) + \log_2 |\det({\bf A}_u)| \notag \\
        &\stackrel{(b)}{=} h({\bf z}_t^u | {\bf y}^{t-1}) - I({\bf z}_t^u; {\bf y}_t | {\bf y}^{t-1}) +R_{\mathrm{exp}}, 
    \end{align}
    where
    \begin{itemize}
        \item[(a)] The translation invariance of differential entropy allows ignoring deterministic control inputs ${\bf u}_t$, and the scaling by ${\bf A}_u$ adds $\log_2 |\det({\bf A}_u)|$.
        \item[(b)] We use the identity $h(X|Y,Z) = h(X|Z) - I(X;Y|Z)$.
    \end{itemize}

    Summing the recursion \eqref{eq:entropy_recursion} from $t=0$ to $T$ yields the terminal entropy:
    \begin{align} \label{eq:terminal_entropy}
        h({\bf z}_{T+1}^u | {\bf y}^T) 
        = h({\bf z}_0^u) + (T+1)R_{\mathrm{exp}} - I({\bf z}_T^u \to {\bf y}^T).
    \end{align}
    
    Applying the condition $\liminf_{T \to \infty} \frac{1}{T+1} I({\bf z}_T^u \to {\bf y}^T) >R_{\mathrm{exp}}$, there exists an $\epsilon > 0$ such that for sufficiently large $T$, $I({\bf z}_T^u \to {\bf y}^T) \ge (T+1)(R_{exp} + \epsilon)$. Substituting this into \eqref{eq:terminal_entropy}:
    \begin{equation} \label{eq:entropy_divergence}
        h({\bf z}_{T+1}^u | {\bf y}^T) \le h({\bf z}_0^u) - (T+1)\epsilon \xrightarrow{T \to \infty} -\infty. \notag
    \end{equation}

We thus establish that the prediction entropy $h({\bf z}_{T+1}^u | {\bf y}^T) \to -\infty$. The posterior entropy relates to the prediction entropy via the mutual information update:
\begin{align} \label{eq:entropy_update}
    h({\bf z}_{T+1}^u | {\bf y}^{T+1}) &= h({\bf z}_{T+1}^u | {\bf y}^T) - I({\bf z}_{T+1}^u; {\bf y}_{T+1} | {\bf y}^T).\notag
\end{align}
Since mutual information is non-negative, the posterior entropy is upper-bounded by the prediction entropy. Consequently, the divergence of the prediction entropy implies the divergence of the posterior entropy:
\begin{equation} \label{eq:posterior_divergence}
    \lim_{T \to \infty} h({\bf z}_{T+1}^u | {\bf y}^{T+1}) = -\infty.\notag
\end{equation}

Next, we relate this entropy to the estimation error. Define the estimation error as ${\bf e}_{T+1} := {\bf z}_{T+1}^u - \hat{{\bf z}}_{T+1}^u$, where $\hat{{\bf z}}_{T+1}^u = \mathbb{E}[{\bf z}_{T+1}^u | {\bf y}^{T+1}]$. By the translation invariance of differential entropy, the entropy of the error equals the posterior entropy of the state:
\begin{equation}
    h({\bf e}_{T+1} | {\bf y}^{T+1}) = h({\bf z}_{T+1}^u - \hat{{\bf z}}_{T+1}^u | {\bf y}^{T+1}) = h({\bf z}_{T+1}^u | {\bf y}^{T+1}).\notag
\end{equation}
Thus, $h({\bf e}_{T+1} | {\bf y}^{T+1}) \to -\infty$.

By Lemma \ref{lemma:log_concavity}, the posterior density of the error ${\bf e}_{T+1}$ becomes strongly log-concave asymptotically. Applying \eqref{eq:entropy_gap} to the error variable ${\bf e}_{T+1}$ with covariance ${\bf \Sigma}_{T+1|T+1}^u$:
\begin{equation}
    \frac{1}{2} \log_2 \left( (2\pi e)^{n_u} \det({\bf \Sigma}_{T+1|T+1}^u) \right) \le h({\bf e}_{T+1} | {\bf y}^{T+1}) + n_u C. \notag
\end{equation}
Combining this with the result that $h({\bf e}_{T+1} | {\bf y}^{T+1}) \to -\infty$, it follows that the Gaussian entropy term must also diverge to $-\infty$:
\begin{equation} \label{eq:det_zero}
    \log_2 \det({\bf \Sigma}_{T+1|T+1}^u) \to -\infty \implies \lim_{T \to \infty} \det({\bf \Sigma}_{T+1|T+1}^u) = 0. \notag
\end{equation}
Hence, the sequence of determinants converges to zero. For the asymptotic analysis, we consider the general time index $t$.

Using Assumption \ref{assump:cond_number}, let $\lambda_{\max, t}$ and $\lambda_{\min, t}$ be the maximum and minimum eigenvalues of the covariance matrix ${\bf \Sigma}_{t|t}^u$. The condition number constraint $\lambda_{\max, t} \le \kappa \lambda_{\min, t}$ implies that the determinant is lower-bounded by the maximum eigenvalue:
\begin{align}
    \det({\bf \Sigma}_{t|t}^u) = \prod_{i=1}^{n_u} \lambda_{i, t} \ge (\lambda_{\min, t})^{n_u} \ge \left( \frac{\lambda_{\max, t}}{\kappa} \right)^{n_u}.\notag
\end{align}
Since $\det({\bf \Sigma}_{t|t}^u) \to 0$ as $t \to \infty$, it necessitates that $\lambda_{\max, t} \to 0$. 
Recall that the mean-square estimation error is the trace of the error covariance matrix. We bound the trace by the dimension $n_u$ times the maximum eigenvalue:
\begin{equation}
    \mathbb{E}[\|{\bf e}_t\|^2] = \text{Tr}({\bf \Sigma}_{t|t}^u) = \sum_{i=1}^{n_u} \lambda_{i, t} \le n_u \lambda_{\max, t} \xrightarrow{t \to \infty} 0.\notag
\end{equation}
This proves the attractivity condition. 
    
Stability follows from convergence: since the sequence converges to 0, it is bounded for $t > T_{\epsilon}$. For the finite initial interval $0 \le t \le T_{\epsilon}$, the error is bounded due to the finite dynamics. Thus, the system is asymptotically mean-square observable.
\end{IEEEproof}

\begin{theorem}[Sufficiency for Asymptotic Stabilizability]
    Under Assumptions 1--3, if $({\bf A}, {\bf B})$ is stabilizable and $\liminf_{T \to \infty} \frac{1}{T+1} I(\mathbf{z}_T^u \to \mathbf{y}^T) >R_{\mathrm{exp}}$ holds, the system is asymptotically mean-square stabilizable.
\end{theorem}
\begin{IEEEproof}
    Theorem \ref{thm-Sufficiency-Observability} guarantees $\lim_{t \to \infty} \mathbb{E}[\|\mathbf{e}_t\|^2] = 0$. Using certainty equivalence $\mathbf{u}_t = \mathbf{K} \hat{\mathbf{z}}_t^u$ with a stabilizing $\mathbf{K}$, the closed-loop dynamics are:
    $\mathbf{z}_{t+1}^u = (\mathbf{A}_u + \mathbf{B}_u \mathbf{K}) \mathbf{z}_t^u - \mathbf{B}_u \mathbf{K} \mathbf{e}_t.$
     Substituting the estimate $\hat{{\bf z}}_t^u = {\bf z}_t^u - {\bf e}_t$ into the system dynamics yields: ${\bf z}_{t+1}^u 
        = {\bf A}_{cl} {\bf z}_t^u + {\bf w}_t,$
    where ${\bf w}_t := -{\bf B}_u {\bf K} {\bf e}_t$ acts as an additive disturbance term driven by the estimation error.
    The closed-loop dynamics represent a stable linear system driven by the input ${\bf w}_t$. Since ${\bf A}_{cl}$ is Schur stable, the system is ISS.
    Hence, the driving term converges to zero in the mean-square sense:
        $\lim_{t \to \infty} \mathbb{E}[\|{\bf w}_t\|^2] \le \|{\bf B}_u {\bf K}\|^2 \lim_{t \to \infty} \mathbb{E}[\|{\bf e}_t\|^2] = 0.$
    For a linear system with a stable system matrix, if the input converges to zero, the state also converges to zero.
\end{IEEEproof}

\section{Conclusion}
This work establishes the fundamental information-theoretic limits for controlling linear systems via nonlinear observations. We proved that stabilizing such systems requires the directed information rate to exceed the intrinsic expansion rate, a necessary condition that becomes sufficient under asymptotic log-concavity. This generalizes classical data-rate theorems beyond the linear-Gaussian paradigm. 

Future work will extend these bounds to stochastic systems with process noise and general nonlinearities. Furthermore, we aim to transition from the fixed-sensing paradigm to an active ``coding'' framework by formulating the sensing mechanism as a variational optimization problem, thereby algorithmically minimizing control error under finite rate constraints.
% \section{Conclusion}
% In this work, we characterized the fundamental information-theoretic limits governing the control and sensing of noiseless linear dynamical systems observed through nonlinear channels. By rigorously deriving necessary conditions, we revealed that the average directed information flow must strictly exceed the intrinsic entropy expansion rate of the unstable dynamics to ensure stabilizability.
% Moreover, we identified asymptotic log-concavity as a key structural property that renders these limits sufficient for asymptotic mean-square stability. This result generalizes classical data-rate theorems beyond the linear-Gaussian paradigm, providing a rigorous theoretical footing for systems where the sensing geometry is the primary bottleneck. 
% Future research will endeavor to extend this framework to stochastic systems with additive process noise. Investigating the universality of the derived bounds across various classes of nonlinearities also represents a promising direction for further work. A particularly compelling next step is to transition from the current fixed-sensing paradigm to an active ``coding'' framework. This entails formulating the sensing and observation mapping as a variational optimization problem, where the encoding policies are no longer governed solely by immutable physical laws but are algorithmically designed to minimize estimation and control error under finite information-rate constraints.

% \newpage

\appendices
\section{Technical Lemmas}

\begin{lemma} \label{lemma:spectral_bounds}
    Let $\mathbf{A}^{-1} = \mathbf{V} \mathbf{J} \mathbf{V}^{-1}$. For a matrix sequence $\mathbf{\Omega}_t = (\mathbf{A}^{-t})^T \mathbf{P} (\mathbf{A}^{-t}) - \sum_{j=0}^{N_t} (\mathbf{A}^{-jL})^T \mathbf{Q}_j (\mathbf{A}^{-jL})$ with $\mathbf{P} \preceq \beta \mathbf{I}, \mathbf{Q}_j \succeq \alpha \mathbf{I}$, the transformed matrix $\tilde{\mathbf{\Omega}}_t := \mathbf{V}^T \mathbf{\Omega}_t \mathbf{V}$ satisfies:
    \begin{equation}
        \tilde{\mathbf{\Omega}}_t \preceq \beta \sigma_{\max}^2(\mathbf{V}) (\mathbf{J}^t)^T \mathbf{J}^t - \alpha \sigma_{\min}^2(\mathbf{V}) \sum_{j=0}^{N_t} (\mathbf{J}^{jL})^T \mathbf{J}^{jL}.\notag
    \end{equation}
\end{lemma}
\begin{IEEEproof}
    Substituting the Jordan decomposition and using singular value bounds $\sigma_{\min}^2(\mathbf{V})\mathbf{I} \preceq \mathbf{V}^T \mathbf{V} \preceq \sigma_{\max}^2(\mathbf{V})\mathbf{I}$ yields the result directly.
\end{IEEEproof}

\begin{lemma} \label{lemma:log_concavity}
    Under Assumptions \ref{assump:obs_hessian} and \ref{assump:prior_hessian}, the Hessian of the log-posterior satisfies $\nabla_{\mathbf{z}_t^u}^2 \log p(\mathbf{z}_t^u | \mathbf{y}^t) \preceq -c \mathbf{I}$ for large $t$.
\end{lemma}
\begin{IEEEproof}
    We define ${\bf H}_{\text{prior}} = \nabla^2_{{\bf z}_0^u} \log p({\bf z}_0^u)$ and ${\bf H}_{\text{obs}, k} := \nabla_{{\bf z}_k^u}^2 \log p({\bf y}_k | {\bf z}_k^u)$.
    The log-posterior Hessian ${\bf H}_t := \nabla_{{\bf z}_t^u}^2 \log p({\bf z}_t^u | {\bf y}^t)$ is thus bounded by:
\begin{align} \label{eq:H_bound_chain}
    {\bf H}_t 
    &= ({\bf A}_u^{-t})^T {\bf H}_{\text{prior}} ({\bf A}_u^{-t}) + \sum_{k=0}^t ({\bf A}_u^{-(t-k)})^T {\bf H}_{\text{obs}, k} ({\bf A}_u^{-(t-k)}) \notag \\
    &\stackrel{(a)}{=} ({\bf A}_u^{-t})^T {\bf H}_{\text{prior}} ({\bf A}_u^{-t}) \notag \\
    &\quad + \sum_{j=0}^{N_t-1} \left( \sum_{k \in \mathcal{T}_j} ({\bf A}_u^{-(t-k)})^T {\bf H}_{\text{obs}, k} ({\bf A}_u^{-(t-k)}) \right) + {\bf R}_t \notag \\
    &\stackrel{(b)}{\preceq} \beta ({\bf A}_u^{-t})^T ({\bf A}_u^{-t}) - \alpha \sum_{j=0}^{N_t-1} ({\bf A}_u^{-jL})^T ({\bf A}_u^{-jL}) + {\bf R}_t,\notag 
\end{align}
where
\begin{itemize}
    \item[(a)] We partition the time horizon $[0:t]$ into $N_t = \lfloor (t+1)/L \rfloor$ blocks of length $L$, indexed by $j \in [0:N_t-1]$. The remainder term is explicitly given by the sum over the initial time steps $\mathcal{T}_{rem} = [0 : t - N_t L]$:
        ${\bf R}_t := \sum_{k=0}^{t - N_t L} ({\bf A}_u^{-(t-k)})^T {\bf H}_{\text{obs}, k} ({\bf A}_u^{-(t-k)}).$
    \item[(b)] We apply Assumption \ref{assump:prior_hessian} (${\bf H}_{\text{prior}} \preceq \beta {\bf I}$) and Assumption \ref{assump:obs_hessian}, where the accumulated Hessian in each block $j$ satisfies the curvature bound $-\alpha ({\bf A}_u^{-jL})^T ({\bf A}_u^{-jL})$.
\end{itemize}

    We employ the Jordan decomposition ${\bf A}_u^{-1} = {\bf V} {\bf J} {\bf V}^{-1}$, where the Jordan matrix ${\bf J}$ is partitioned as: $\mathbf{J} = \begin{bmatrix} \mathbf{J}_{<1} & \mathbf{0} \\ \mathbf{0} & \mathbf{J}_{=1} \end{bmatrix}.$
    Applying Lemma \ref{lemma:spectral_bounds} , the transformed Hessian $\tilde{{\bf H}}_t := {\bf V}^T {\bf H}_t {\bf V}$ satisfies:
    $\tilde{\mathbf{H}}_t \preceq \beta \sigma_{\max}^2(\mathbf{V}) (\mathbf{J}^t)^T \mathbf{J}^t 
    + C_{\mathrm{rem}} \sum_{k \in \mathcal{T}_{\mathrm{rem}}} (\mathbf{J}^{t-k})^T (\mathbf{J}^{t-k}) 
    - \alpha \sigma_{\min}^2(\mathbf{V}) \sum_{j=0}^{N_t-1} (\mathbf{J}^{jL})^T (\mathbf{J}^{jL}). $
    For strictly unstable modes ($\rho(\mathbf{J}_{<1}) < 1$), the negative term dominates asymptotically as the prior vanishes. For marginally unstable modes ($\rho(\mathbf{J}_{=1}) = 1$), the cumulative sum of the negative term (order $O(t^{2K-1})$) asymptotically dominates the positive prior term (order $O(t^{2K-2})$). Thus, $\mathbf{H}_t$ becomes negative definite.
\end{IEEEproof}

\section*{Acknowledgement}
This work was supported in part by the National Science and Technology Major Projects of China under Grant 2025ZD1302000, and the National Natural Science Foundation of China (NSFC) under Grant 62522107.

\bibliographystyle{IEEEtran}
\bibliography{ref}

\end{document}